\documentclass[12pt]{article}
\usepackage{graphicx}
\newlength{\extraspace}
\newlength{\extraspaces}
\catcode`\@=11
\def\numberbysection{\@addtoreset{equation}{section}
\def\theequation{\arabic{section}.\arabic{equation}}}

\begin{document}
\addtolength{\baselineskip}{.7mm}
\thispagestyle{empty}
\begin{flushright}
TIT-HEP-448 \\
KEK-TH-696 \\
{\tt hep-ph/0005231} \\
May, 2000 
\end{flushright}
\vspace{2mm}
\begin{center}
{\large{\bf Natural Mass Hierarchy of $Z$ Boson and Scalar Top
\\ in No-Scale Supergravity 
}}

\vspace{5mm}
{\sc Yuichi Chikira}\footnote{
\tt e-mail: ychikira@th.phys.titech.ac.jp}   \\[2mm]
{\it Department of Physics, Tokyo Institute of Technology \\
Oh-okayama, Meguro, Tokyo 152-0033, Japan} \\[2mm]
and \\[2mm]
{\sc Yukihiro Mimura}\footnote{
\tt e-mail: mimura@ccthmail.kek.jp}  \\[2mm]
{\it Theory Group, KEK, Oho 1-1, Tsukuba, Ibaraki 305-0801, Japan} \\[5mm]
{\bf Abstract}\\[5mm]
{\parbox{14cm}{\hspace{5mm}
%
%
%
A study has shown that
a `no-scale' model makes a hierarchy between the scalar top mass 
and the Z boson mass naturally.
The supersymmetry breaking parameters are constrained by 
flavor changing neutral currents in the minimal supersymmetric standard model.
One solution of the problem is that the gaugino mass is the 
only source of supersymmetry breaking parameters at the Planck scale.
However, in such a scenario, we need a cancellation between the Higgs 
mass parameters under the minimization condition of the Higgs potential.
We insist that there is no such cancellation in the no-scale model,
and that the no-scale model provides a prediction of the scalar top mass 
and the lightest Higgs mass.
The lightest Higgs mass is predicted to be $m_H = 110 \pm 5 $ GeV.

}}
\end{center}
\vfill
\newpage
\setcounter{section}{0}
\setcounter{equation}{0}
\setcounter{footnote}{0}
\def\theequation{\arabic{section}.\arabic{equation}}
%
%
%
\pagebreak[3]
\addtocounter{section}{1}
\setcounter{equation}{0}
\setcounter{subsection}{0}
\setcounter{footnote}{0}
\begin{center}
{\large {\bf \thesection. Introduction}}
\end{center}
\nopagebreak
\nopagebreak
\hspace{3mm}

Supersymmetric theories now stand as the most promising candidates
for a unified theory beyond the standard model \cite{Nilles}.
Accurate data remarkably favor the supersymmetric grand 
unified theory (GUT) over any non-supersymmetric theory~\cite{Amaldi}.
%
Supersymmetry helps to resolve the gauge hierarchy problem 
\cite{Sakai}.
In non-supersymmetric standard models, because the squared Higgs mass 
receives a quadratic divergent correction radiatively,
we cannot explain the hierarchy between the weak scale
and the grand unified scale naturally.
Supersymmetry removes the quadratic divergences and provides 
a framework for naturally explaining the widely separated hierarchy.

Under those contexts,
the idea of radiative breaking of the electroweak symmetry~\cite{Inoue}
is very popular.
It is very attractive to explain the breaking of electroweak symmetry
through large logarithms between the Planck (or GUT) scale and the weak scale.
The radiative corrections drive an up-type Higgs mass-squared 
parameter negative for a large top Yukawa coupling, 
and thus the electroweak symmetry breaks down.
The radiative symmetry breaking mechanism has consequences
for the supersymmetric particle spectrum, and provides important 
constraints on the particle spectrum.

These constraints also provides a slight puzzle.
In the radiative breaking mechanism, the $Z$ boson mass is 
related to the supersymmetry breaking parameters.
We thus believe that supersymmetric particles
are not very heavy compared with the $Z$ boson.
However, the experimental lower bounds for the supersymmetric 
particle masses are becoming larger day by day,
and it seems that we must require a fine tuning
between the parameters in the Higgs potential \cite{Giudice}.

It is well known that flavor changing neutral currents (FCNC) 
make important constraints on the supersymmetry breaking scalar masses \cite{Nilles,FCNC}.
We require that the scalar quark eigenmasses have degeneracy\footnote{
   In addition to the degeneracy scenario,
   there is an alignment scenario, in which
   the scalar quark eigenvectors should be strongly aligned
   with those of the quark eigenvectors.}
of a few percent when the scalar masses are on the order of O(100) GeV.
%
%
One solution concerning the scalar quark mass degeneracy
is to consider the type of minimal gaugino mediation \cite{MGM}.
%
%
Namely,
the gaugino mass is almost the only source for supersymmetry breaking
at the Planck scale.
The supersymmetry breaking parameters which have flavor indices
are sufficiently small compared to the gaugino mass at the Planck scale, 
and become sufficiently large due to renormalization group flow at the low energy scale.
Though this scenario is very attractive, 
such large gaugino masses cause the fine tuning described above to
the Higgs potential.
Are there any mechanisms in which fine tuning is not 
required, even if the gaugino mass is large?

In this paper, 
we insist that the `no-scale' supergravity model \cite{Lahanas}
does not require any fine tuning of the Higgs potential.
The no-scale models are very suitable for the scenario of 
minimal gaugino mediation.
We consider the supersymmetric
particle spectrum in no-scale models, where 
the magnitude of the supersymmetry breaking 
parameters is also determined radiatively.
We can investigate the theoretical upper bounds 
in no-scale models,
and can judge the bounds at 
near future colliders.
Especially, we insist that the no-scale models suggest
a natural mass hierarchy between the $Z$ boson mass and the supersymmetry breaking masses.

The organization of this paper is as follows.
In Section 2, we review 
unnatural tuning in the Higgs potential
in Minimal Supersymmetric Standard Model (MSSM).
In Section 3, we review no-scale supergravity models.
In Section 4, we explain how to calculate the particle spectrum
in our framework.
In Section 5, we show the results for the particle spectrum
and discuss its bounds.
Finally, we conclude in Section 6 with a summary of our results.

%
%
%
\pagebreak[3]
\addtocounter{section}{1}
\setcounter{equation}{0}
\setcounter{subsection}{0}
\setcounter{footnote}{0}
\begin{center}
{\large {\bf \thesection. Unnatural Tuning in $Z$ Boson Mass}}
\end{center}
\nopagebreak
\nopagebreak
%
%

The tree level neutral Higgs potential in MSSM is given by
\begin{equation}
V^{(0)} = m_1^2 |H_d^0|^2 + m_2^2 |H_u^0|^2
    - (m_3^2 H_d^0 H_u^0 + c.c.)
    + \frac{g^2 + g'^2}8 (|H_d^0|^2 - |H_u^0|^2)^2.
\label{tree_potential}
\end{equation}
The Higgs mass parameters, $m_1^2$ and $m_2^2$, are
\begin{equation}
m_1^2 = m_{H_d}^2 + \mu^2, \qquad m_2^2 = m_{H_u}^2 + \mu^2,
\end{equation}
where $m_{H_d}^2$ and $m_{H_u}^2$ are the soft supersymmetry breaking
mass squared for the Higgs bosons,
and $\mu$ is the so-called Higgsino mass in the supersymmetric '$\mu$-term'.
We denote the vacuum expectation values (VEVs) for $H_d^0$ and $H_u^0$
as $v_d$ and $v_u$, respectively.

We require that electroweak symmetry breaks down, and then  
find the minimization conditions of the potential at the tree level,
\begin{equation}
\frac{M_Z^2}2 = - \mu^2 
                  + \frac{ m_{H_d}^2 - m_{H_u}^2 \tan^2 \beta }
                         { \tan^2 \beta -1 },
\label{minimization_1}
\end{equation}
\begin{equation}
\sin 2\beta = \frac{2 m_3^2}{m_1^2 + m_2^2},
\label{minimization_2} 
\end{equation}
where $\tan \beta = v_u/v_d$.
We require that the $Z$ boson mass, $M_Z$, is equal to
91 GeV and that $\tan \beta$ is not so close at 1,
phenomenologically.
We should mention here that a relation like 
Eq.(\ref{minimization_1}) is usually satisfied even 
in non-minimal models.

Radiative electroweak symmetry breaking occurs
because $m_{H_u}^2$ is driven negative due to a large top
Yukawa coupling in its renormalization group flow.
It is well-known \cite{King} that the heavy gluino mass causes a 
weird cancellation
among supersymmetry breaking Higgs mass squareds and $\mu$
in the $Z$ boson mass formula (\ref{minimization_1}).

\begin{figure}[tbp]
\begin{center}
\includegraphics[width=8cm,clip]{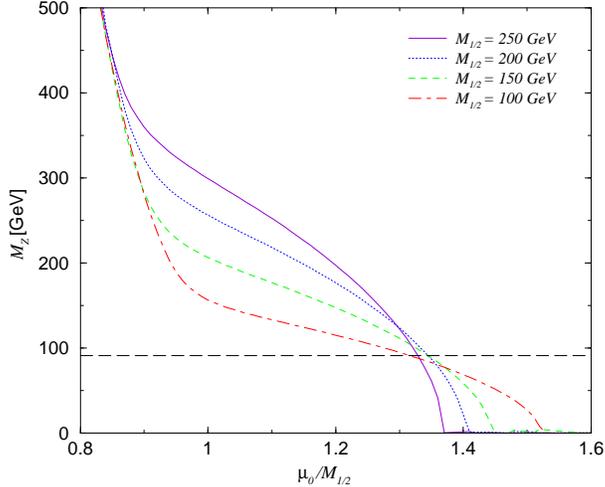}
\end{center}
\caption{We show the $Z$ boson mass as a function of $\mu_0/M_{1/2}$
         for various gaugino masses. In this figure, we set $m_0$ and $A_0$
         to be zero.
         We choose the $B_0$ parameter so that $\tan \beta=10$ 
         at the point $M_Z = 91$ GeV.
         On the left side of this figure, $\tan \beta$ becomes close to 1,
         and the Higgs potential is destabilized.}
\label{fig:1}
\end{figure}

Let us clarify such an unnatural cancellation.
The `free' dimensionful parameters for MSSM is
\begin{equation}
\{ m_0^2, M_{1/2}, A_0, B_0, \mu_0 \}.
\end{equation}
These parameters are introduced at the Planck scale.\footnote{
Since we would not like to consider the Physics beyond GUT,
those parameters are given at the GUT scale later.}
The main contribution for negative $m_{H_u}^2$ is
not the original supersymmetry breaking scalar mass squared $m_0^2$
at the GUT scale,
but the gluino mass, $M_{\tilde g}$ \cite{Moroi,Giudice}.
The scalar mass-squared $m_0^2$ is insensitive to negative $m_{H_u}^2$
in the regime $m_0 \sim O(100)$ GeV.
Since the most sensitive parameters
for the Higgs-mass squared
are the gaugino mass, $M_{1/2}$, and $\mu$
among those parameters,
the other three parameters are equal to zero for the time being.
We show Fig.\ref{fig:1}\footnote{
To plot the figure, we consider the 1-loop corrected
scalar potential (\ref{1-loop_potential}).}
in which we plot the $Z$ boson mass
as a function of $\mu/M_{1/2}$ for a given $M_{1/2}$.
%
This figure provides us with the problem:
why is the parameter $\mu$ limited to within a narrow range
for an appropriate electroweak symmetry breaking,
even when we solve the $\mu$-problem\footnote{
We have a problem which is so called the $\mu$-problem;
why is the supersymmetric parameter $\mu$
of the same order as the supersymmetry breaking parameters.
}?
Besides, the parameter $\mu$ should be the
right edge value in the figure for the allowed region,
if the gaugino mass, $M_{1/2}$, is larger than 200 GeV.
In fact, $M_{1/2}$ should be larger than about 200 GeV
in the minimal gaugino mediation noted in Section 1.


Let us view fine tuning in Eq.(\ref{minimization_1}) from another point of view.
Here, we suppose that $\tan \beta$ is sufficiently large ($\tan\beta > 3$) only for simplicity.
Then, the $Z$ boson mass is written by
\begin{equation}
M_Z^2 = -2 (\mu^2 + m_{H_u}^2) = -2 m_2^2.
\end{equation}
Since parameters $\mu$ and $m_{H_u}^2$ depend on the scale $Q$,
we should know the scale where we require tuning between 
$\mu^2$ and $m_{H_u}^2$. 
The scale is the one where 1-loop corrected potential becomes small.
We denote the scale as $Q_{\tilde t}$,
since it is nearly equal to the mass of scalar top quarks.
Then, the physical $Z$ boson mass is approximated by
\begin{equation}
M_Z^2 \sim M_Z^2(Q_{\tilde t}),
\end{equation}
where
\begin{equation}
M_Z^2(Q) \equiv  -2 m_2^2(Q).
\end{equation}
We define the scale $Q_0$ where $M_Z^2(Q)$ vanish.
The $Q_0$ is the scale where electroweak symmetry breaks down at the tree level.
Expanding $M_Z^2(Q)$ by $\ln Q$ around scale $Q_0$, we obtain
\begin{equation}
M_Z^2 \sim 2 \ln (\frac{Q_0}{Q_{\tilde t}}) \frac{d m_2^2}{d \ln Q} (Q_0).
\label{Z_expand}
\end{equation}
{}From this point of view, 
the fine tuning in the $Z$ boson mass is translated into 
the tuning between  
the scalar quark mass scale $Q_{\tilde t}$ and 
the scale $Q_0$.
As stated in Ref.\cite{Gamberini}, 
electroweak symmetry breaks down only when
the scale $Q_{\tilde t}$ is less than $Q_0$.
This fact means the same thing as saying that the parameter $\mu$ should be the right
edge in Fig.\ref{fig:1} when the gaugino mass becomes greater.

There are many implications about naturalness in the literature.
In Ref.\cite{Moroi}, it is pointed out that
a heavy scalar mass $m_0$ (which masses are
of the order of 1 TeV) relaxes the fine tuning.
Ref.\cite{King} suggests that a less fine-tuned
model should be selected as a scenario candidate for supersymmetry breaking.
It is preferred that the gaugino mass is not unified at the GUT scale
(e.g. D-brane model)
in the reference.
In this literature, 
we feel that the fine tuning in the Higgs potential is not dispelled.
Are there any models in which cancellation occurs naturally?

In this paper, we suggest that we have already had a model 
which can explain the heavy gluino mass naturally without any 
fine tuning.
This model is no-scale supergravity.
In folklore, it is said that
more severe fine tuning is required in the no-scale supergravity
model rather than in ordinary models.
We believe, however, that this interpretation is not correct.
To see this, we give a brief review of no-scale supergravity
in the next section.

%
%
%
\pagebreak[3]
\addtocounter{section}{1}
\setcounter{equation}{0}
\setcounter{subsection}{0}
\setcounter{footnote}{0}
\begin{center}
{\large {\bf \thesection. No-Scale Supergravity}}
\end{center}
\nopagebreak
\nopagebreak
%
%

In this section, we briefly review the no-scale supergravity,
and we consider the natural mass hierarchy between the $Z$ boson
and the scalar top in the no-scale model.

In the hidden sector model \cite{hidden},
we separate fields into two sectors, 
which are a visible sector and a hidden sector.
The observable fields (quarks, leptons and Higgs fields) 
are involved in the visible sector.
The hidden fields, which break supersymmetry, exist in the
hidden sector, and couple with the
observable fields through only gravitational interaction.
The $F$ terms of the hidden fields have VEVs due to the
dynamics in only the hidden sector in ordinary hidden sector models.
In other words, the scale of supersymmetry breaking is
determined with no relation to our visible sector.
However, 
it is possible that a scalar potential for the hidden sector fields
is flat at the tree level, and that the VEVs of 
the hidden fields determine radiatively accompanied with
visible sector dynamics.
Such theories are called no-scale supergravity \cite{Lahanas}. 

Let us see how the gravitino mass is determined in the no-scale model.
Using the minimization conditions (\ref{minimization_1}) and 
(\ref{minimization_2}),
we obtain the tree level MSSM scalar potential at the minimal point,
\begin{equation}
V_{\rm mim}^{(0)} = - \frac1{2 (g^2 + g'^2)} M_Z^4 \cos^2 2\beta.
\end{equation}
Since the $Z$ boson mass is proportional to the gravitino mass\footnote{
          We assume that the dimensionful parameters
          are proportional to the gravitino mass. 
          See Appendix.},
the potential involving the hidden sector is unbounded from below. 
However, there exists a 1-loop corrected scalar potential, 
\begin{equation}
V^{(1)}=
\frac1{64\pi^2} \sum_J (-1)^{2J} (2J+1) m_J^4 (\ln \frac{m_J^2}{Q^2} - \frac32),
\label{1-loop_potential}
\end{equation}
in $\overline{\rm DR}$ scheme \cite{Coleman}.
As a result, the scalar potential is stabilized if Str $M^4 > 0$ \cite{Kounnas},
and the gravitino mass is determined dynamically.

We emphasize here that the gravitino mass is not independent 
on the visible sector parameter, namely $\mu$, in the no-scale models.
The naturalness argument in the no-scale model
differs from arguments in the ordinary ones
due to such a dependence.

To confirm the natural hierarchy between the $Z$ boson mass 
and the supersymmetry breaking masses,
we will overview the minimization with respect to gravitino mass \cite{Lahanas}.
Since the total scalar potential does not depend on the renormalization point,
we can evaluate the potential at a scale where the 1-loop corrected potential vanishes,
\begin{equation}
V^{(1)} (v_u,v_d;Q) = 0.
\end{equation}
This scale is approximately the mass scale of the scalar top quarks ($Q_{\tilde t}$),
\begin{equation}
Q_{\tilde t} \equiv (m_{\tilde t_1} m_{\tilde t_2})^{1/2}.
\end{equation}
We can then find the minimal value of the effective scalar potential \cite{Lahanas},
\begin{equation}
V_{\rm min}  
\sim -C Q_{\tilde t}^4 \left(\ln \frac{Q_{\tilde t}^2}{Q_0^2}\right)^2,
\end{equation}
where $Q_0$ is the scale where electroweak symmetry breaks down at the tree level,
and $C$ is a constant.
In the no-scale model, $Q_{\tilde t}$ is determined, by which 
the $V_{\rm min}$ is minimized.
By minimizing $V_{\rm min}$ for $Q_{\tilde t}$,
we find that the scale $Q_{\tilde t}$ is determined as
\begin{equation}
\ln \frac{Q_0^2}{Q_{\tilde t}^2} = 1.
\end{equation}
It is important that $Q_{\tilde t}$ is
very close to $Q_0$,
\begin{equation}
Q_{\tilde t} = Q_0/e^{1/2}.
\end{equation}

Substituting it into Eq.(\ref{Z_expand}), 
we find the $Z$ boson mass formula in the no-scale model
as follows for large $\tan \beta$:
\begin{equation}
M_Z^2 \sim \frac{d}{d \ln Q} m_2^2.
\label{Z_no-scale}
\end{equation}
The 1-loop renormalization group equations (RGEs) for $m_{H_u}^2$ and $\mu^2$
are
\begin{equation}
\frac{d}{d\ln Q} m_{H_u}^2 = \frac1{2\pi}
                 [ 3 \alpha_t (m_{\tilde t_3}^2 + m_{\tilde t}^2
                               + m_{H_u}^2 + A_t^2)
                  - (\alpha' M_1^2 + 3 \alpha_2 M_2^2) ],
\end{equation}
\begin{equation}
\frac{d}{d\ln Q} \mu^2  = \frac1{2\pi}
           [ 3 \alpha_t + 3 \alpha_b + \alpha_\tau
             - (\alpha' + 3 \alpha_2 ) ] \mu^2,
\end{equation}
where $\alpha_t = Y_t^2/4 \pi$ and $Y_t$ is a top Yukawa coupling.
It turns out that the
$Z$ boson mass is determined hierarchically compared to 
the supersymmetry breaking masses,
and that the hierarchy is characterized by 1-loop factor $3\alpha_t/2\pi$.
This fact is what we insist on in this paper.

Eq.(\ref{Z_no-scale}) is easily extended in the case of a general $\tan \beta$.
Expanding Eq.(\ref{minimization_1}) by $\ln Q$ around $Q_0$,
we obtain the following $Z$ boson mass formula at the tree level:
\begin{equation}
M_Z^2 \cos^2 2\beta \sim  \dot m_1^2 \cos^2\beta
                        + \dot m_2^2 \sin^2\beta
                        - \dot m_3^2 \sin 2\beta,
\end{equation}
where $\dot m_i^2 = d m_i^2 /d \ln Q$.
This relation will be tested in the future experiments.

When introducing this formula, we neglect the derivative
of the 1-loop corrected potential with respect to Higgs VEVs
in the $Z$ boson mass formula.
Since it is complicate to write down the derivative,
we calculate the 1-loop corrected relation numerically.
We show how we obtain our numerical results in the next section.

%
%
%
\pagebreak[3]
\addtocounter{section}{1}
\setcounter{equation}{0}
\setcounter{subsection}{0}
\setcounter{footnote}{0}
\begin{center}
{\large {\bf \thesection. Methods}}
\end{center}
\nopagebreak
\nopagebreak
%
%

We concentrate on the following effective scalar potential with a
Higgs VEVs independent shift:
\begin{equation}
V_{\rm eff}(v_u,v_d) =
 V^{(0)}(v_u,v_d;Q) 
 + V^{(1)}(v_u,v_d;Q) - V^{(1)}(v_u,v_d=0;Q).
\label{effective_potential}
\end{equation}
This potential is independent of the renormalization point, $Q$,
at the 1-loop level schematically \cite{Casas}.
In the expression, $V^{(0)}$ is a tree level potential (\ref{tree_potential}) 
and $V^{(1)}$ is a 1-loop correction (\ref{1-loop_potential}) of the potential.

\begin{figure}[tbp]
\begin{center}
\includegraphics[width=8cm,clip]{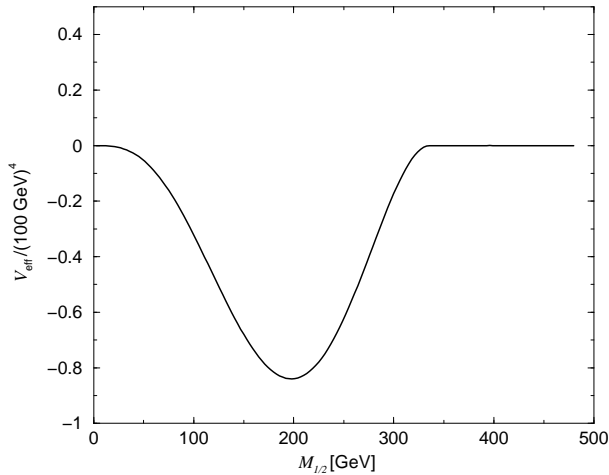}
\end{center}
\caption{Effective potential Eq.(\ref{effective_potential})
         minimized by the Higgs vacuum expectation values, $v_u$ and $v_d$.}
\label{fig:2}
\end{figure}

At first, we show an effective potential which is 
minimized by Higgs VEVs $v_d$ and $v_u$ (Fig.\ref{fig:2}).
This figure is drawn in the minimal case, 
where $m_0^2 = 0$ and $A_0=B_0=0$.
The horizontal axis is for the gaugino mass.
We can see that there is a minimum with respect to the gaugino mass.


Since the gaugino mass is the most sensitive parameter for
the supersymmetry breaking Higgs mass squared,
we normalize the following dimensionful 
parameters of MSSM:
\begin{equation}
\{ m_0^2, M_{1/2}, A_0, B_0, \mu_0 \},
\end{equation}
divided by the gaugino mass $M_{1/2}$,
and we adopt the following four dimensionless parameters:
\begin{equation}
\{ \hat m_0^2, \hat A_0, \hat B_0, \hat \mu_0 \}
\end{equation}
as parameters for no-scale models.
The {\it hat} denotes that the parameters are
normalized by the gaugino mass (squared).
The gaugino mass is determined in minimizing the potential 
if we fix the hatted parameters\footnote{
The gravitino mass, $m_{3/2}$, is also a parameter of the model,
but is still a free parameter, since the proportional coefficient,
$m_{3/2}/M_{1/2}$, is not determined in the model.}.
The freedom for $\hat \mu_0$ is consumed 
when the $Z$ boson mass is fixed as 91 GeV.
If we fix $\tan \beta$, $\hat B_0$ is consumed
and  
the remaining free parameters are only $\hat m_0$ and $\hat A_0$.

To show our numerical analysis,
we evolve the supersymmetry breaking parameters
with the full two-loop RGEs \cite{RGE2}.

Since the potential does not ideally depend on the renormalization
point, we may choose any scale.
Nevertheless, we minimize the potential (\ref{effective_potential})
near to the scale where the electroweak symmetry breaks down 
{\it at the tree level} to fix our aim.
This is because we must consider the threshold effect for
supersymmetric particles, for instance, scalar quarks and gluinos.
Therefore, we adopt our method as follows.
Firstly, we minimize the effective potential with respect to the
Higgs VEVs and the gaugino mass at the scale above those supersymmetric
particle masses.
After minimization, we include one-loop threshold
corrections from supersymmetric particles \cite{threshold},
and fix the physical quantities.
We take as inputs 
$\alpha_{\rm em}^{-1} (M_Z) = 127.9$,
$\sin^2 \theta_W (M_Z)_{\overline{\rm MS}}= 0.2309$
and
$M_Z = 91.2$ GeV.

The strong gauge coupling has a discrepancy
between the prediction from GUT and the
experimental measurement.
The value of the strong gauge coupling, $\alpha_3$,
is predicted to be $\alpha_3 (M_Z) = 0.13$ in GUT,
while in the experimental measurement, 
$\alpha_3 (M_Z) = 0.119$.
We adopt the experimental value for the strong gauge coupling.
The resulting particle spectra have a great dependence
upon the strong gauge coupling and top Yukawa coupling.
For smaller gauge coupling, the supersymmetric particles
become heavier.
This is mainly because the top Yukawa coupling at GUT scale is
bigger for the smaller gauge coupling.

We assume that the gaugino masses are unified at the GUT scale\footnote{
The GUT scale is defined as the scale where $\alpha_1 = \alpha_2$,
which are the gauge coupling constants.
}.
We also assume the universality of $m_0^2$ and $A_0$
for their flavor and matter indices at the GUT scale for simplicity.

In our calculation, the bottom quark mass and the tau lepton mass are
fixed as $m_b (M_Z) = 3.0$ GeV and
$m_\tau (M_Z) = 1.7$ GeV.
The results we will show later have little dependence on the
bottom and tau masses.
The top quark pole mass is fixed as $M_t = 174$ GeV.
The 1-loop relationship between the pole mass and tree level
mass $Y_t v_u$ is given by 
$M_t = Y_t v_u (1+ 5\alpha_3/3\pi)$ in the
$\overline{\rm DR}$ scheme.

%
%
%
\pagebreak[3]
\addtocounter{section}{1}
\setcounter{equation}{0}
\setcounter{subsection}{0}
\setcounter{footnote}{0}
\begin{center}
{\large {\bf \thesection. Numerical Results}}
\end{center}
\nopagebreak
\nopagebreak
%
%


It is convenient that we present the
RGE solution by the following parameterization \cite{Giudice}.
The dimensionful parameters at low energy are written
using the GUT scale parameters.

First of all, the up-type Higgs mass squared, $m_2^2$, is written as
\begin{equation}
m_2^2 = 1.0 \mu_0^2 - 0.05 m_0^2 - 1.75 M_{1/2}^2
                  -0.34 M_{1/2} A_0 - 0.10 A_0^2
\label{RGsolution:1}
\end{equation}
in the case of $\tan \beta = 10$.
The mass of the $Z$ boson is $M_Z^2 \sim -2 m_2^2$.
It is easy to see that we require fine tuning 
between $\mu_0$ and $M_{1/2}$, if the gaugino is much heavier 
than $Z$ boson.
It is worth noting that the coefficient of $m_0^2$ is very small
in the RGE solution in the expression of $m_2^2$.
This is because the 'focus-point scale' for $m_{H_u}^2$ is of the
order of 100 GeV \cite{Moroi}\footnote{
In the reference \cite{Moroi}, it seems that 
the coefficient of $m_0^2$ is of opposite sign to ours.
In our calculation, the sign is reversed when we take the top quark pole mass 
as $M_t = 172$ GeV.
}.

In contrast, the RGE solution for $d m_2^2/d \ln Q $
is
\begin{equation}
\frac{d m_2^2}{d \ln Q} 
= 0.015 \mu_0^2 + 0.026 m_0^2 + 0.245 M_{1/2}^2
 - 0.011 M_{1/2} A_0 - 0.004 A_0^2
\label{RGsolution:2}
\end{equation}
when we take $\tan \beta$ = 10.
The $Z$ boson mass in the no-scale model is
$M_Z^2 \sim dm_2^2/ d \ln Q$.
We can easily see that the tuning required above is not necessary 
in the no-scale model.

Eqs.(\ref{RGsolution:1}) and (\ref{RGsolution:2}) 
are important for understanding our numerical results 
qualitatively,
such that $-2 m_2^2 \sim 
dm_2^2/d\ln Q = (100-110 {\rm GeV})^2$.
The numerical value (100-110 GeV) for $M_Z$ is caused by
the 1-loop corrected potential \cite{Arnowitt}.

We will give the numerical results of minimizing the
potential, including the 1-loop corrected potential.
In the following figures, the sign of the $\mu$ parameter is 
positive in the notation used in Eq.(\ref{LR_mixing}).

\begin{figure}[tbp]
\begin{center}
\includegraphics[width=8cm,clip]{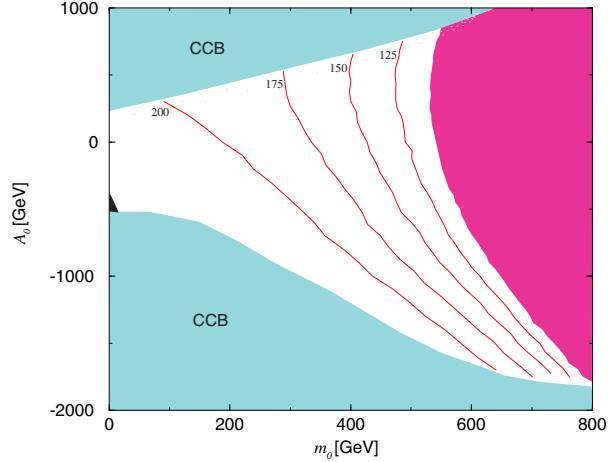}
\end{center}
\caption{Contour plot for the gaugino mass as a function of $m_0$ 
         and $A_0$ in the case of $\tan\beta =10$.}
\label{fig:3}
\end{figure}

\begin{figure}[tbp]
\begin{center}
\includegraphics[width=8cm,clip]{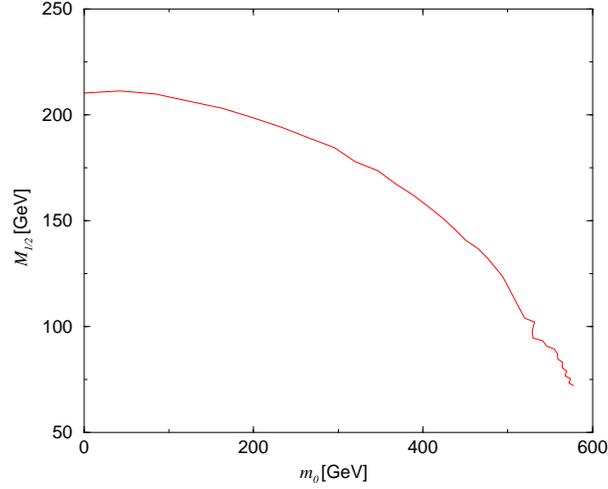}
\end{center}
\caption{$m_0$-$M_{1/2}$ plot at $A_0 = 0$ in the case of $\tan\beta=10$.}
\label{fig:4}
\end{figure}

In Fig.\ref{fig:3},
we show a contour plot for the gaugino mass, $M_{1/2}$,
as a function of $m_0$ and $A_0$ in the case of $\tan \beta=10$.
The shaded area on the right side of the figure is excluded for 
the condition $M_{1/2} > 100$ GeV,
which means that the lightest chargino is heavier than 85 GeV.
The upper and lower areas are excluded for charge and color breaking (CCB),
namely:
\begin{eqnarray}
A_t^2 &>& 3 (m_{\tilde q_3}^2 + m_{\tilde t}^2 + m_{H_u}^2),  \nonumber \\
A_b^2 &>& 3 (m_{\tilde q_3}^2 + m_{\tilde b}^2 + m_{H_d}^2),            \\
A_\tau^2 &>& 3 (m_{\tilde \ell_3}^2 + m_{\tilde \tau}^2 + m_{H_d}^2).  \nonumber 
\end{eqnarray}
In the black area at $m_0 \sim 0$ and $A_0 \sim -500$ GeV,
the right-hand scalar tau is lighter than the lightest neutralino,
which is not preferred for neutralino LSP.
We note that 
the $m_0$ and $A_0$ are tuned in the lower right region in the figure.
We prefer a small $m_0$ because of FCNC constraints.
Therefore, we do not regard the large $m_0$ region.
In Fig.\ref{fig:4},
we show $m_0$-$M_{1/2}$ plot at $A_0 = 0$.
We can qualitatively understand 
its elliptic shape from Eq. (\ref{RGsolution:2}).

\begin{figure}[tbp]
\begin{center}
\includegraphics[width=8cm,clip]{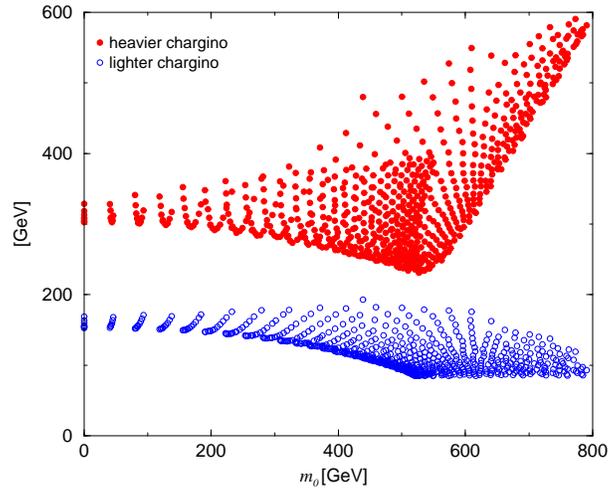}
\end{center}
\caption{Chargino masses.
         The heavier and lighter chargino masses are approximately $\mu$ and
         the wino mass respectively.
	 We cut the lighter chargino mass, which is smaller than 85 GeV.
         }
\label{fig:5}
\end{figure}

\begin{figure}[tbp]
\begin{center}
\includegraphics[width=8cm,clip]{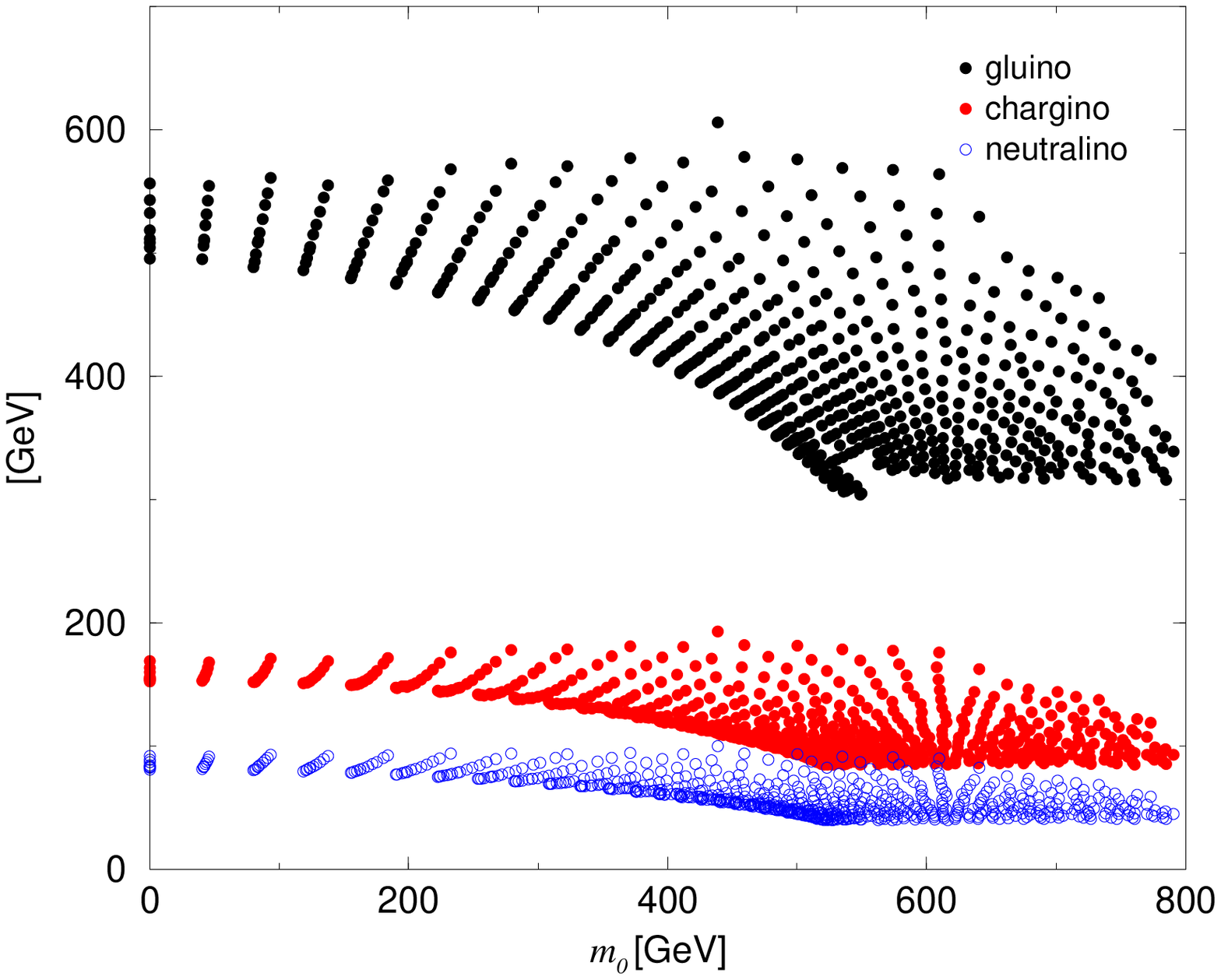}
\end{center}
\caption{Gluino mass, lightest chargino mass and lightest neutralino mass.
 	 The 1-loop correction for gluino mass is included.
         The lighter chargino mass, which is smaller than 85 GeV, has been cut.}
\label{fig:6}
\end{figure}

Fig.\ref{fig:5}
shows the chargino masses as a function of $m_0$ for various $A_0$.
We remark that
the dots are plotted every 0.2 interval for $\hat m_0$ (not $m_0$),
and 0.5 interval for $\hat A_0$, thus
the density of the dots is not related to the probability of the parameters.
This remark is also applied in the figure below. 
In Fig.\ref{fig:6},
we show the gluino mass, lightest chargino mass and lightest neutralino mass 
as a function of $m_0$ for various $A_0$ in the same way as in Fig.\ref{fig:5}.

\begin{figure}[tbp]
\begin{center}
\includegraphics[width=8cm,clip]{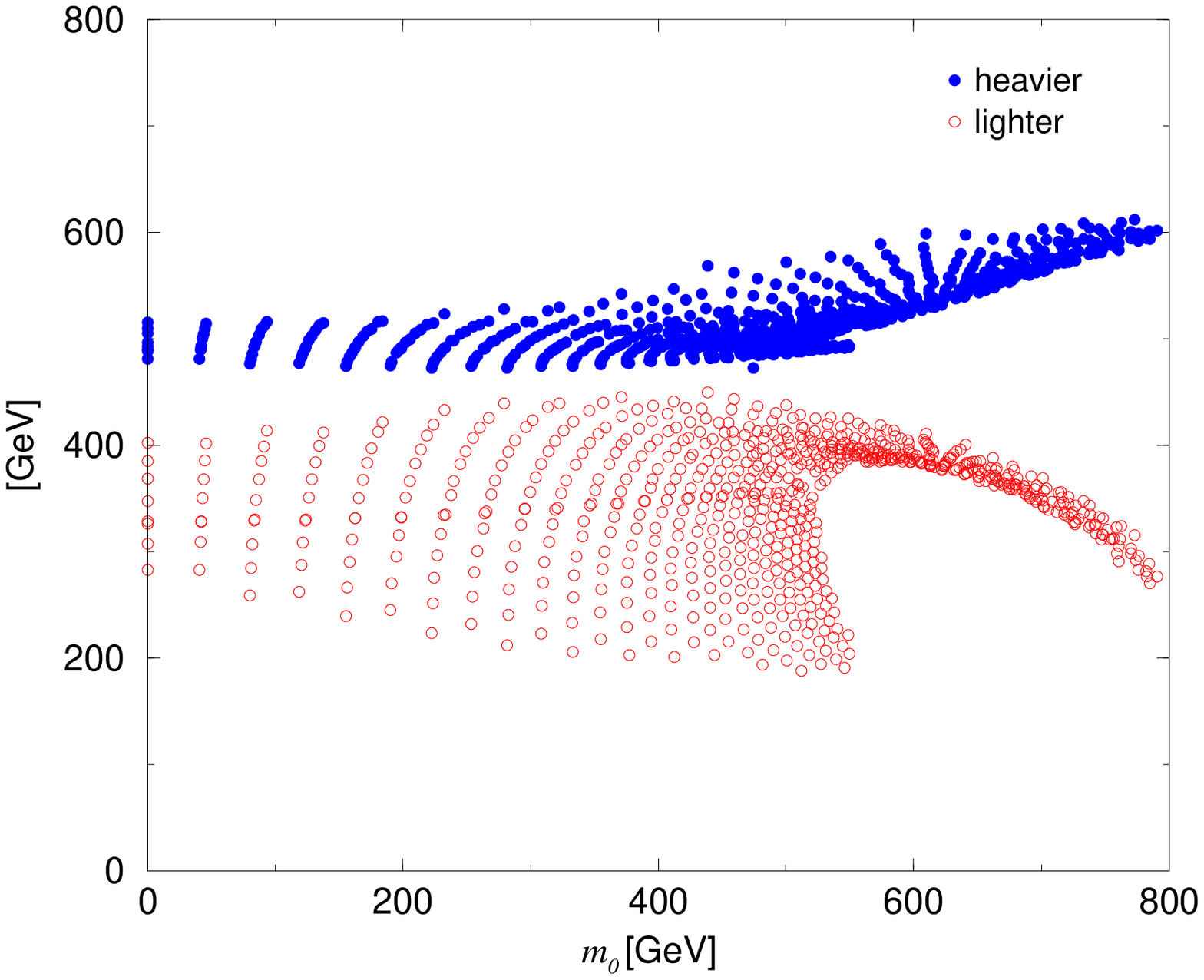}
\end{center}
\caption{Scalar top masses.
         The scalar top masses are determined up to left-right mixing.}
\label{fig:7}
\end{figure}

The important prediction for the no-scale model is
that the scalar top masses are almost determined independently 
of the scalar mass, $m_0$.
The scalar top masses are plotted in Fig.\ref{fig:7}.

\begin{figure}[tbp]
\begin{center}
\includegraphics[width=8cm,clip]{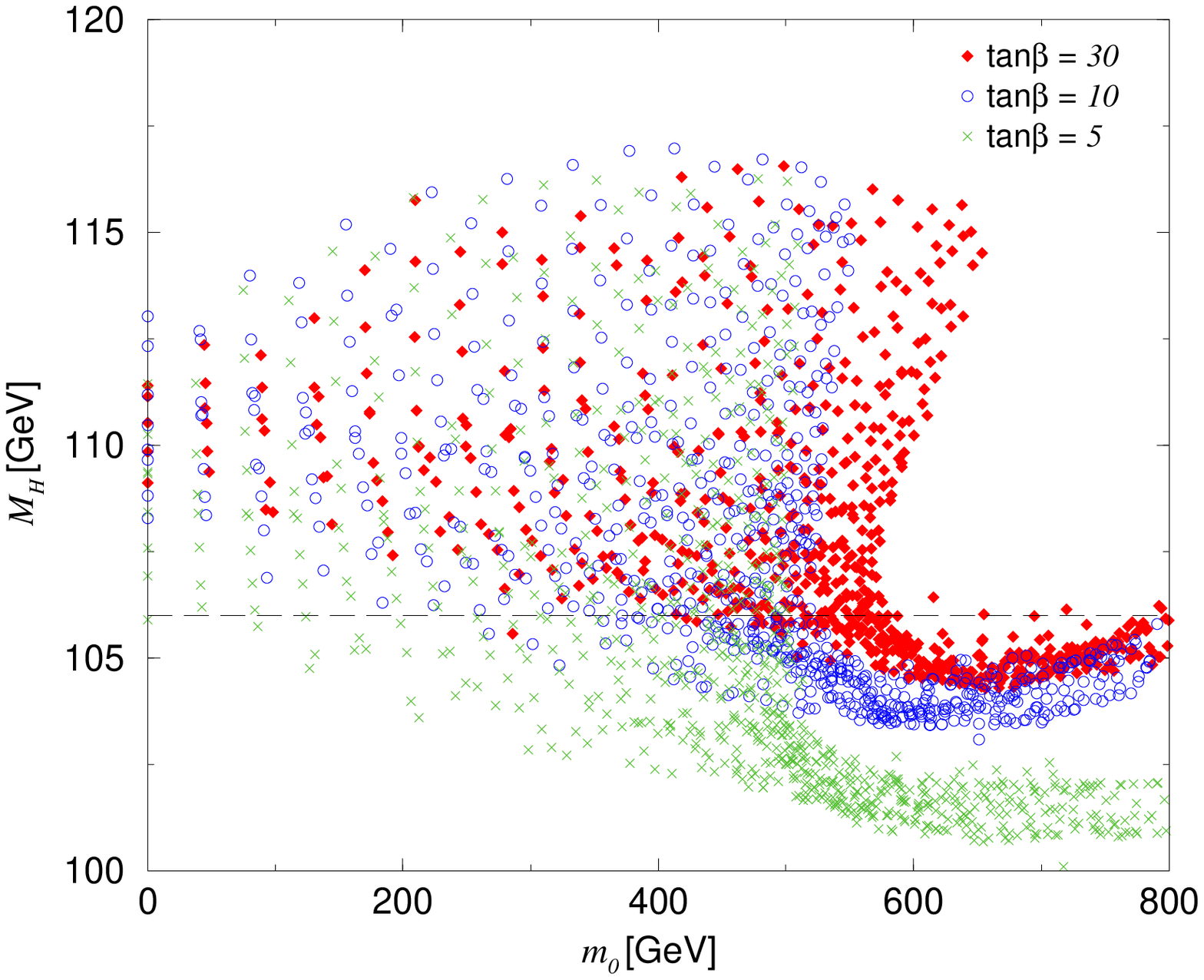}
\end{center}
\caption{Lightest Higgs mass for $\tan\beta=$ 5,10,30.
The dots are plotted every 0.2 interval for $\hat m_0$ (not $m_0$),
and 0.5 interval for $\hat A_0$.
The density of the dots is not related to the probability of the parameters.
We line the recent LEPII bound on the non-observation of 
$e^+e^- \rightarrow ZH$ \cite{Marciano} for one's information.
}
\label{fig:8}
\end{figure}

In supersymmetric models, the lightest Higgs mass is bounded by $M_Z$ 
at the tree level. However, this upper bound is corrected by the 
1-loop potential \cite{Okada}.
Since the scalar top masses are almost determined, 
the lightest Higgs mass is also predictable in the no-scale model.
We plot the lightest Higgs mass for $\tan \beta =$ 5, 10, 30 in Fig.\ref{fig:8}.
It is important that the lightest Higgs mass is $110 \pm 5$ GeV for small $m_0$.
The small $m_0$ is favored for FCNC constraints.
In calculating the lightest Higgs mass, we adopt the 2-loop approximate formula 
for the mass in Ref.\cite{Higgs_mass}.

%
%
%
\pagebreak[3]
\addtocounter{section}{1}
\setcounter{equation}{0}
\setcounter{subsection}{0}
\setcounter{footnote}{0}
\begin{center}
{\large {\bf \thesection. Discussion}}
\end{center}
\nopagebreak
\nopagebreak
%
%

\begin{figure}[tbp]
\begin{center}
\includegraphics[width=8cm,clip]{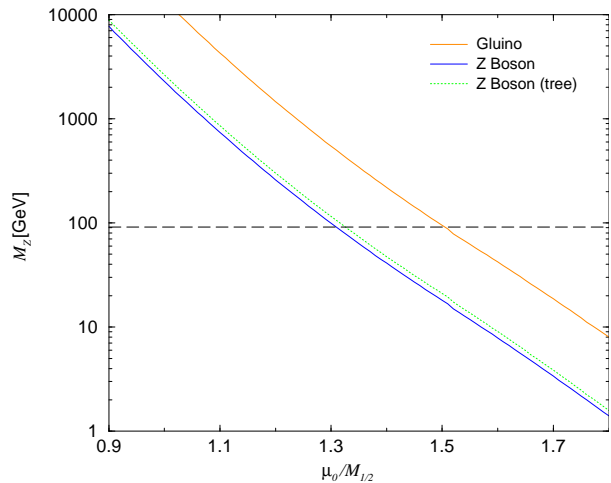}
\end{center}
\caption{$Z$ boson mass and gluino mass as a function of $\mu_0/M_{1/2}$
         in the case of the no-scale model.
	 Their mass ratio is approximately constant to $\mu_0/M_{1/2}$.
         We also plot the tree level $Z$ boson mass formula, Eq.(\ref{Z_no-scale}).}
\label{fig:9}
\end{figure}

In order to see our insist visibly,
we show figure (Fig.\ref{fig:9}) in the corresponding plot to Fig.\ref{fig:1}.
Again, we set parameter $m_0^2$ and $A_0$ to be zero.
We choose the $B_0$ parameter so as to be $\tan\beta=$10 at the point 
$M_Z=$91 GeV\footnote{
Incidentally, $\tan \beta$ is just 10 when the $B_0$ parameter is equal to zero.
}.
We plot the $Z$ boson mass as a function of 
$\mu_0/M_{1/2} (= \hat \mu_0 )$.
There is no weird constraint for the parameter $\hat \mu_0$
for electroweak symmetry breaking,
contrary to the case in Fig.1.
Therefore, the model-building God can create the MSSM parameters
without considering whether electroweak symmetry can break down 
at low energy.
Our $Z$ boson mass (91 GeV) does
not lie on a special point, contrary to the ordinary case.

The following quantity \cite{Barbieri}
is usually used for measuring the sensitivity of the $Z$ boson mass
for variations in parameter $a$,
\begin{equation}
\Delta_a = \left| \frac{\partial \ln M_Z^2}{\partial \ln a} \right|.
\end{equation}
The value of $\Delta_{\hat \mu_0}$ is also large in no-scale model.
Namely,
the $Z$ boson mass is sensitive to the parameter $\hat \mu_0$.
However, the large value of $\Delta_{\hat \mu_0}$ in the no-scale model
does not cause any fine-tuning problem, contrary to ordinary models.
It is just an event that a value of the $Z$ boson mass is selected.

Some people may say that it is also just an event in other 
supersymmetry breaking scenarios.
This opinion is obviously true.
However, the predictive ability in the no-scale model is completely 
different.
%
The subtractive tuning in the ordinary model does not 
have any predictive power.
The supersymmetry breaking mass scale may be of the order of 10 TeV
in the ordinary model.
On the other hand, 
we do not require any subtractive tuning in the no-scale model,
and predict that 
all of the supersymmetric particles (except gravitino)
appear below about 500-600 GeV.
Especially, we can judge the no-scale model 
when we search the Higgs boson or gauginos in the near future.
This predictive ability is our motivation concerning the no-scale model.
%
%
For theoretical physicists, it is important to search predictive models.
%
To say more, it is important that 
we recognize that the fine-tuning in the Higgs potential 
may impose the no-scale supergravity,
and we consider predictions of the no-scale models.
This is a process of Physics to access the unknown world.

{\bf Note added:}
While completing this paper, we received a paper by R. Barbieri
and A. Strumia \cite{Strumia} which also considers that
the electroweak breaking scale becomes related to the supersymmetry breaking scale
by a loop factor in a similar way to us.

%
%
\pagebreak[3]
\addtocounter{section}{1}
\setcounter{equation}{0}
\setcounter{subsection}{0}
\setcounter{footnote}{0}
\begin{center}
{\large {\bf  Acknowledgments
}}
\end{center}
\nopagebreak
\medskip
\nopagebreak
\hspace{3mm}
Y.M. would like to thank to N. Okada for the discussion of the
no-scale supergravity.
This work was supported by 
JSPS Research Fellowships for Young Scientists.

\renewcommand{\thesection}{A}
\section{Notation and Convention}\label{appendix}
\setcounter{equation}{0}
\renewcommand{\theequation}{A.\arabic{equation}}

The superpotential of minimal supersymmetric standard model (MSSM)
is presented as
\begin{equation}
W = Y_u Q \cdot H_u U^c + Y_d H_d \cdot Q D^c + Y_e H_d \cdot L E^c
    + \mu H_d \cdot H_u,
\end{equation}
where the SU(2) inner product is defined as
\begin{equation}
H_d \cdot H_u \equiv H_d^T \epsilon H_u, \qquad 
\epsilon = \left(
                 \begin{array}{cc}
                 0 & 1 \\
                 -1 & 0
                 \end{array}
           \right).
\end{equation}
Here, $Q$, $U^c$, $D^c$, $L$, $E^c$ are matter chiral superfields,
and $H_u$ and $H_d$ are Higgs doublets.

We denote the soft supersymmetry breaking terms as
\begin{eqnarray}
V_{soft} &=&
             m_{H_d}^2 |H_d|^2 + m_{H_u}^2 |H_u|^2 \nonumber \\
&& + m_{\tilde q}^2 \tilde q \tilde q^\dagger
   + m_{\tilde u}^2 \tilde u_R \tilde u_R^\dagger
   + m_{\tilde d}^2 \tilde d_R \tilde d_R^\dagger
   + m_{\tilde \ell}^2 \tilde \ell \tilde \ell^\dagger
   + m_{\tilde e}^2 \tilde e_R \tilde e_R^\dagger \nonumber \\
&& + (A_u Y_u \tilde q \cdot H_u \tilde u_R^c
   + A_d Y_d H_d \cdot \tilde q \tilde d_R^c
   + A_e Y_e H_d \cdot \tilde \ell \tilde e_R^c + h.c.) \nonumber \\
&& + ( B \mu H_d \cdot H_u + h.c. )
\end{eqnarray}

To clarify our notation, we present the left-right component 
in the scalar top quark mass matrix and chargino mass matrix
in the following.
The left-right mixing is 
\begin{equation}
(A_t + \mu \cot \beta) m_t. 
\label{LR_mixing}
\end{equation}
The chargino mass matrix is presented as
\begin{equation}
M_{\chi^+} = \left(
             \begin{array}{cc}
             M_2 & \sqrt2 M_W \cos \beta \\
             \sqrt2 M_W \sin \beta & -\mu 
             \end{array}
             \right).
\end{equation}

The supergravity theories are
given by the K\"ahler potential $K$, the superpotential $W$
and the gauge kinetic function $f$.
The scalar potential is given in supergravity as
\begin{equation}
V = e^K [ g^{ij^*} (D_i W) (D_{j^*} W^*) - 3 W W^* ].
\end{equation}
Using the K\"ahler transformation $G = K + \log W + \log W^*$,
we obtain 
\begin{equation}
V = e^G [ G^i G_i - 3 ].
\end{equation}

The no-scale K\"ahler potential \cite{Lahanas} is written as
\begin{equation}
G = - 3 \ln (T+\bar T - h(\phi_i^*,\phi_i)) + \ln | W(\phi_i) |^2,
\label{Kahler}
\end{equation}
where $T$ is a moduli field and $\phi$ are fields in the visible sector.
The function $h$ is a K\"ahler potential for the visible fields.
The scalar potential is thus
\begin{equation}
V = \frac{3|W|^2}{(T+ \bar T -h(\phi_i^*,\phi_i))^2}
\left| \frac{\partial W}{\partial \phi_i} \right|^2.
\end{equation}
If the global supersymmetric conditions, $\partial W/\partial \phi_i = 0$,
are satisfied, the scalar potential for $T$ is flat and the gravitino mass,
$m_{3/2} = e^{G/2}$,
is not determined.

Expanding the K\"ahler potential with respect to visible fields, $Q$,
we write the K\"ahler potential \cite{Kaplunovsky} as
\begin{equation}
K = \hat{K}(T,T^*) + \tilde{K}_{ij^*}(T,T^*) Q^i Q^{j^*}
  + \frac12 ( H_{ij}(T,T^*) Q^i Q^j + h.c.) + \cdots,
\end{equation}
where the $T^\alpha$'s are hidden sector fields (Dilaton and Moduli).
The superpotential is given by
\begin{equation}
W = \hat W + \tilde \mu_{ij} Q^i Q^j 
              + \tilde Y_{ijk} Q^i Q^j Q^k + \cdots.
\end{equation}

Substituting VEV into $T$,
we obtain the effective superpotential in the flat limit,
\begin{equation}
W_{\rm eff} (Q) = \mu_{ij} Q^i Q^j + Y_{ijk} Q^i Q^j Q^k.
\end{equation}
The parameters $\mu$ and $Y$ are written as \cite{Kaplunovsky}
\begin{eqnarray}
\mu_{ij} = e^{\hat K/2} \frac{\hat W^*}{|\hat W|} \tilde \mu_{ij}
           + m_{3/2} (H_{ij} - G^{\alpha^*} \partial_{\alpha^*} H_{ij}),
\end{eqnarray}
\begin{equation}
Y_{ijk} =  e^{\hat K/2} \frac{\hat W^*}{|\hat W|} \tilde Y_{ijk}.
\end{equation}
In order to solve the $\mu$-problem, we often set $\tilde \mu$ 
to be zero \cite{Masiero}.
Then, the $\mu$ parameter in the flat limit is proportional to the gravitino mass.

The gaugino mass is given by the gauge kinetic function, $f_a$, as
\begin{equation}
M_a = \frac12 m_{3/2} G^{\alpha} \partial_\alpha f_a.
\end{equation}
The supersymmetry breaking scalar mass squared is
\begin{equation}
m^2_{ij^*} = m_{3/2}^2 
          (\tilde K_{ij^*} 
           - G^\alpha G^{\beta^*} R_{\alpha \beta^* i j^*}).
\end{equation}
Those two parameters are proportional to the gravitino mass (squared).

The $A$ and $B$ parameters are complicated, and are not necessarily
proportional to the gravitino mass.
In this paper, we suppose that $A$ and $B$ are proportional to the gravitino mass
for simplicity.

We note that the parameters $m_0$, $A$ and $B$ are zero at the Planck scale
in a strict no-scale model, which is given by the K\"ahler potential (\ref{Kahler}).
In this paper, we loosen the boundary condition. 



\vspace{5mm}
%
\newcommand{\NP}[1]{{\it Nucl.\ Phys.\ }{\bf #1}}
\newcommand{\PL}[1]{{\it Phys.\ Lett.\ }{\bf #1}}
\newcommand{\CMP}[1]{{\it Commun.\ Math.\ Phys.\ }{\bf #1}}
\newcommand{\MPL}[1]{{\it Mod.\ Phys.\ Lett.\ }{\bf #1}}
\newcommand{\IJMP}[1]{{\it Int.\ J. Mod.\ Phys.\ }{\bf #1}}
\newcommand{\PRP}[1]{{\it Phys.\ Rep.\ }{\bf #1}}
\newcommand{\PR}[1]{{\it Phys.\ Rev.\ }{\bf #1}}
\newcommand{\PRL}[1]{{\it Phys.\ Rev.\ Lett.\ }{\bf #1}}
\newcommand{\PTP}[1]{{\it Prog.\ Theor.\ Phys.\ }{\bf #1}}
\newcommand{\PTPS}[1]{{\it Prog.\ Theor.\ Phys.\ Suppl.\ }{\bf #1}}
\newcommand{\AP}[1]{{\it Ann.\ Phys.\ }{\bf #1}}
\newcommand{\ZP}[1]{{\it Zeit.\ f.\ Phys.\ }{\bf #1}}
\end{document}